# Intrinsic Phonon Bands in High-Quality Monolayer *T'* Molybdenum Ditelluride

*Shao-Yu Chen[†], Carl H. Naylor[§], Thomas Goldstein[†], A.T. Charlie Johnson[§], and Jun Yan[†*]*

[†]Department of Physics, University of Massachusetts, Amherst, Massachusetts 01003, United States

[§]Department of Physics and Astronomy, University of Pennsylvania, Philadelphia, Pennsylvania 19104, United States

[*]Corresponding Author: Jun Yan.    Tel:    (413)545-0853          Fax:    (413)545-1691
E-mail: yan@physics.umass.edu

**Abstract**

The topologically nontrivial and chemically functional distorted octahedral ($T'$) transition metal dichalcogenides (TMDCs) are a type of layered semimetal that has attracted significant recent attention. However, the properties of monolayer (1L) $T'$-TMDC, a fundamental unit of the system, is still largely unknown due to rapid sample degradation in air. Here we report that well-protected 1L CVD $T'$-MoTe$_2$ exhibits sharp and robust intrinsic Raman bands, with intensities about one order of magnitude stronger than those from bulk $T'$-MoTe$_2$. The high-quality samples enabled us to reveal the set of all nine even-parity zone-center optical phonons, providing reliable fingerprints for the previously elusive crystal. By performing light polarization and crystal orientation resolved scattering analysis, we can effectively distinguish the intrinsic modes from Te-metalloid-like modes A (~122 cm$^{-1}$) and B (~141 cm$^{-1}$) which are related to the sample degradation. Our studies offer a powerful non-destructive method for assessing sample quality and for monitoring sample degradation *in situ*, representing a solid advance in understanding the fundamental properties of 1L-$T'$-TMDCs.

Molybdenum and tungsten based transition metal dichalcogenides (TMDCs) are a class of layered materials hosting versatile phases and polymorphs that exhibit fascinating physical, chemical and topological properties.[1–5] Hexagonal (*H*) TMDCs such as $(Mo,W)(S,Se)_2$ are semiconductors with bandgaps ranging from visible to infrared, and distorted octahedral (*T'*) TMDCs such as $WTe_2$ and $MoTe_2$ are semimetals that possess nontrivial electronic band topologies.[4–6] As a layered material system these compounds and their intercalated products are being actively investigated for solid lubricant, catalyst, and super-capacitor applications.[7–10] In the fundamental monolayer (1L) limit, *H*-TMDCs exhibit direct bandgap with coupled spin and valley physics,[11] and have been widely investigated for electrical, optoelectronic and valleytronic applications.[12–14] In contrast 1L distorted octahedral (*T'*) TMDCs, predicted to be two dimensional (2D) topological insulators,[5] are much less studied experimentally.

It is of both fundamental and practical interest to develop a thorough understanding of *T'*-TMDC at the monolayer level. Since the prediction of the quantum spin Hall (QSH) effect in *T'* monolayer, several works have studied thin films[15,16] (~10 nm thick) of *T'*-$MoTe_2$ and *T'*-$WTe_2$ to assess if the QSH gap is positive or negative, which is important for developing topological quantum computing devices using this system. Monolayer *T'*-TMDC can serve as a bottom-up starting point for understanding multi-layered and intercalated *T'*-TMDCs, which have much higher electrical conductivity than their *H* counterparts, and could be more useful for solid state battery electrodes, electrochemical capacitors, and hydrogen evolution reactions.[9,10]

As it turns out 1L-*T'*-TMDC is relatively challenging to work with due to rapid sample degradation in air.[17,18] For this reason, several recent optical and electrical studies of the material system [15–17] are limited to multi-layers; and despite recent efforts,[19–21] a thorough Raman characterization of 1L-*T'*-TMDC is still lacking. Also, the sample degradation has led to confusion in distinguishing Raman bands intrinsic to *T'*-TMDC from those due to degradation products.[22,23] In this paper, we show that well-protected 1L-$T'$-MoTe$_2$ exhibits sharp and robust intrinsic Raman bands, with intensities about one order of magnitude stronger than those from bulk $T'$-MoTe$_2$. The high-quality samples enabled us to reveal the set of *all* nine even-parity zone-center optical phonons, representing a complete Raman characterization of 1L-$T'$-TMDC. We also discovered that less well protected samples show coexistence of phonon spectra due to $T'$-MoTe$_2$ and its degradation products; the two types of Raman features exhibit drastically different symmetry properties, and can be unambiguously distinguished by light-polarization and crystal-angle resolved Raman tensor analysis. These experimental findings provide key insights into fundamental lattice dynamics of the crystal, and offer a powerful non-destructive means for assessing sample quality and for monitoring sample degradation *in situ*, paving way to topological transport investigations and to further understanding of the chemical and physical properties of $T'$-TMDCs. Our studies also clear up recent confusions on investigations of strain[22] and laser[23] induced $H$ to $T'$ phase transition in MoTe$_2$ crystals, and show that care needs to be taken in distinguishing Raman features due to the phase transition and those due to sample degradation.

**Results and Discussion**

The 1L-$T'$-MoTe$_2$ studied in this paper is grown by chemical vapor deposition (CVD) on 300 nm SiO$_2$/Si substrate. The growth is achieved at 700°C with rapid thermal quench (See Methods for more details).[24] Figure 1a shows the optical image of a typical sample. The CVD grown 1L-$T'$-MoTe$_2$ flakes have a bamboo-leaf like shape; this elongated shape reflects the underlying atomic scale in-plane anisotropy due to spontaneous formation of zigzag Mo atomic chains as illustrated in Figure 1e. The step-height from substrate to the monolayer is about 0.8 nm as shown by atomic force microscope (AFM) measurements in Figure 1b, which is consistent with the 0.7-0.8 nm per layer thickness from previous studies.[25] The crystal grows preferentially along the zigzag chains, which we denote as the *a*-axis. This can be seen from the transmission electron microscope (TEM) diffraction pattern (Figure 1d) of a sample transferred onto a holey-carbon TEM grid (Figure 1c): the rectangular reciprocal lattice is in accord with the rectangular real space unit cell (light blue rectangle in Figure 1e), from which we determine the in-plane unit vector lengths of the crystal to be $a$ = 3.42 Å, $b$ = 6.34 Å. For later description of the lattice vibrations we also show in Figure 1e the position of a mirror symmetry plane in the crystal (red horizontal line, perpendicular to the zigzag Mo atomic chain).

With the knowledge of the zigzag atomic chain direction, we proceed with Raman scattering measurements on the monolayer $T'$ crystal, paying special attention to the angle between directions of the $MoTe_2$ $a$-axis and the light polarization. To protect the $T'$-$MoTe_2$, we cover the sample with CVD graphene (see Methods), whose Raman features[26] have much higher energy than those in $T'$-$MoTe_2$. The sample is excited by a linearly polarized Nd:YAG laser at 532 nm, with the laser power kept below 200 $\mu$W. We denote the angle between the incident light polarization and the crystal $a$-axis as $\theta$ (as shown in Figure 2a), which can be continuously adjusted by rotating the laser polarization with a half waveplate. The scattered photons are directed through a set of half waveplate and linear polarizer in the collection path to select the co-linear (HH) or cross-linear (HV) component, which is subsequently dispersed by a triple spectrometer and detected with a liquid-nitrogen-cooled CCD camera.[27] Throughout the light-polarization and crystal-angle resolved Raman measurements, the sample is mounted inside an optical cryostat pumped down to $3 \times 10^{-6}$ torr.

Figure 2b compares typical room temperature spectra of 1L and bulk $T'$-$MoTe_2$ sample in the HH configuration at $\theta = 0°$ using the same excitation power. The overall spectral shapes are similar below 200 cm$^{-1}$ with the bulk Raman peaks slightly red-shifted due to interlayer coupling. Interestingly in contrast to studies of 1L-$T'$-$WTe_2$,[19–21] the Raman peaks of our monolayer samples are intense and sharp, suggesting that our $T'$-$MoTe_2$ monolayers are of high crystalline quality.

In Figure 2c we selectively display the Raman spectra of 1L-$T'$-$MoTe_2$ in four different scattering configurations of HV and HH at $\theta = 0°$ and 45°; more detailed intensity angular dependences are plotted in Figure 3c. We observe in total nine zone-center optical phonons, including six '$m$-modes' ($m^{85}$, $m^{113}$, $m^{128}$, $m^{164}$, $m^{253}$ and $m^{270}$) and three '$z$-modes' ($z^{92}$, $z^{102}$ and $z^{190}$), where '$z$' stands for the zigzag Mo atomic chain (black arrow in Figure 1e), and '$m$' stands for the mirror plane perpendicular to the zigzag chain (red line in Figure 1e). The reason for this classification becomes clear in Figure 3 where we have plotted detailed angular dependence of Raman intensity in HH and HV together with the atomic displacements for corresponding phonons. We observe that for the three '$z$-modes' the vibrations are along the zigzag Mo atomic chain ($a$-axis), and that for the six '$m$-modes' the vibrations are in the mirror plane perpendicular to the zigzag chain direction ($b$-$c$ plane).

The assignment and classification for the nine Raman bands can be understood from symmetry considerations. In Figure 3a we plot the top and side views of monolayer $T'$-MoTe$_2$ together with its unit cell and symmetry operations (the symbols are consistent with the International Table for Crystallography).[28] The primitive unit cell, indicated as the shaded area, contains six atoms, resulting in 18 Brillouin zone center ($\Gamma$) phonons whose atomic displacements are illustrated in Figure 3c. Symmetry operations of the crystal include, in addition to translations, identity ($E$), inversion ($i$), mirror reflection ($m$), and a screw axis along the zigzag Mo chain ($2_1^z$) that form the $C_{2h}$ group. Similar to the bulk $T'$-TMDC crystal with monoclinic stacking, 1L $T'$-MoTe$_2$ is centrosymmetric with its inversion centers inside the atomic layer (yellow dots in Figure 3a). For this reason, half of the 18 zone-center vibrations have even-parity and the other half have odd-parity. Since in crystals with inversion centers the even Raman-active and the odd infrared-active modes are mutually exclusive, the nine modes we observe are in fact, the maximum number of zone-center optical phonons that are allowed to appear in the Raman spectra.

Another important symmetry of 1L-$T'$-MoTe$_2$ is the mirror reflection operation (the mirrors are parallel to the $b$-$c$ plane perpendicular to the zigzag direction; see red lines in Figure 1e and also Figure 3a). This mirror plane operation is shared by many $T'$-TMDC polytypes[29] and thus provides a generic way to categorize phonons as: *m*-modes, atomic vibrations in the mirror plane; and *z*-modes, atomic vibrations perpendicular to the mirror plane along the zigzag atomic chain. Because vibrations in a plane have twice degrees of freedom as compared with vibrations along a line, there are 12 *m*-modes and 6 *z*-modes among the 18 zone center phonons in 1L-*T'*-MoTe$_2$. Taking into account the inversion symmetry as discussed above, half of *m*- and half of *z*-modes have even parity, while the other half have odd parity. As a result, in monolayer *T'*-MoTe$_2$, there are 3 Raman active *z*-modes and 6 Raman active *m*-modes. Our spectra shown in the Figure 2c display all 9 different modes, consistent with symmetry analysis and representing a complete Raman characterization of the zone-center optical phonons.

The mode assignment is further substantiated by the angular dependence of Raman intensity which is sensitive to the symmetry of phonon modes. The $C_{2h}$ group has four irreducible representations (see Figure 3b) and the zone-center phonon representations are given by $3A_u + 6B_u + 6A_g + 3B_g$. Raman active *z*-modes are odd under *m* operation ($m = -1$) and are even under inversion operation ($i = 1$), resulting in the symmetry $B_g$. Raman active *m*-modes are even under both mirror reflection and inversion operation, resulting in the symmetry $A_g$. The theoretical angular dependence of Raman-active mode intensity is given by $I = A|\langle \epsilon_i | R^T \cdot \mathcal{R} \cdot R | \epsilon_o \rangle|^2$, where $A$ is a constant, $\epsilon_i$ and $\epsilon_o$ are polarizations of the incident and outgoing light respectively, $\mathcal{R}$ is the effective Raman tensor linked to different symmetry (for *m*-modes $\mathcal{R}_{A_g} = \begin{bmatrix} d & 0 & f \\ 0 & e & 0 \\ f & 0 & h \end{bmatrix}$, and for *z*-modes $\mathcal{R}_{B_g} = \begin{bmatrix} 0 & g & 0 \\ g & 0 & j \\ 0 & j & 0 \end{bmatrix}$),[30] $R$ is the in-plane (*x*-*y* plane) rotation matrix to determine the angle between light polarization and *a*-axis given by $R = R_z = \begin{bmatrix} \cos(\theta) & -\sin(\theta) & 0 \\ \sin(\theta) & \cos(\theta) & 0 \\ 0 & 0 & 1 \end{bmatrix}$. We let $\epsilon_i = \begin{bmatrix} 1 \\ 0 \\ 0 \end{bmatrix}$, then $\epsilon_o$ equals $\begin{bmatrix} 0 \\ 1 \\ 0 \end{bmatrix}$ in HV and $\begin{bmatrix} 1 \\ 0 \\ 0 \end{bmatrix}$ in HH. It is then straightforward to calculate the Raman intensity angular dependence for the *z* and *m* modes: $I^z_{HV}(\theta) \propto \cos^2(2\theta)$, $I^z_{HH}(\theta) \propto \sin^2(2\theta)$, $I^m_{HV}(\theta) \propto \sin^2(2\theta)$, and $I^m_{HH}(\theta) \propto |d \cos^2\theta + e \sin^2\theta|^2$. This explains why the three *z*-modes in Figure 3c have the same angular dependence, while the six *m*-modes have the

same angular dependence in HV but not in HH, since the latter depends on the phase difference between $d$ and $e$ in the effective Raman tensor $\mathfrak{R}_{A_g}$ which are different for different lattice vibrations, and exhibits 2-fold, instead of 4-fold, symmetric patterns, reflecting the anisotropic phonon-specific interactions of $T'$-TMDC crystal.[29,31,32] We note that the observed angular dependences here in the monolayer are exactly the same as those for the $z$-modes and $m$-modes in bulk $T'$-TMDC with either monoclinic or orthorhombic stacking,[29] attesting our previous statement that this classification of lattice vibrations in $T'$-TMDC crystals is very general.

The Raman data in Figures 2 & 3 are taken on flakes with the best quality. We have found that there is a correlation between optical contrast and sample quality: as can be seen in Figure 1a, the atomic flakes on the Si/SiO$_2$ substrate have similar shape but different darkness, and in general darker samples are of better quality. The variations of optical contrast and sample quality in certain areas on the silicon chip are likely due to non-uniform passivation from either water residue left during the transfer process (see Methods) or incomplete protection from CVD graphene that we use to cover the $T'$-MoTe$_2$ sample. There could be voids and cracks in the CVD graphene that were either innate from growth or created during the transfer process. In Figure 4 we show typical Raman spectra of four flakes with different optical contrast. Sample S1 is has 'good' contrast and the Raman features are similar to the data in Figure 2, indicating that it is well passivated; sample S4 is poorly passivated and has 'poor' contrast whose 1L-$T'$-MoTe$_2$ Raman features are almost gone, and instead, two intense new peaks (A to B) between 100 and 150 cm$^{-1}$ show up; meanwhile samples S2 and S3 are in-between with S2 displaying coexistence of Raman features from 1L-$T'$-MoTe$_2$ and from sample degradation. We note that our observation of the correlation between optical contrast and sample degradation is consistent with other investigations of air sensitive 2D materials in literature.[20,33,34]

To understand the origin of these new peaks, we measured Raman spectra of tellurium powder (Sigma-Aldrich 99.997%); see the upper panel of Figure 4b. The Raman spectra of Te powder show two prominent modes $A_1$ and $E$ (following the naming convention in crystalline Te),[35,36] corresponding to the Te breathing vibration in the basal plane and asymmetric stretching vibration along the *c*-axis, respectively. These are similar to A and B modes we observed in sample S4, suggesting that Te metalloids are likely a by-product of MoTe$_2$ degradation. Note that there are some small energy differences between the Raman peaks in Te powder and degraded MoTe$_2$: peak A has slightly higher energy in degraded MoTe$_2$ and redshifts from 128 cm$^{-1}$ in S2 to 122 cm$^{-1}$ in S4, approaching the 121 cm$^{-1}$ peak in Te powder, suggesting different Te cluster sizes in S2 to S4.[37,38] We also note that this peak could overlap with the $m^{128}$ mode in 1L-*T'*-MoTe$_2$; however the former is typically much more intense. Meanwhile the Te-like side peak B (~142 cm$^{-1}$) is useful for monitoring sample quality, since 1L-*T'*-MoTe$_2$ is spectrally clean between 130 and 155 cm$^{-1}$.

To further confirm that the Raman peaks A and B are indeed due to sample degradation we collected polarization and angular dependent data similar to those in Figures 2&3. As can be seen in Figure 4c, the spectra taken on the degraded sample at $\theta = 0°$ and 45° are almost identical in either HH or HV, in drastic contrast to intrinsic 1L-*T'*-MoTe$_2$ peaks in Fig.2c. The detailed angular dependence of peaks A and B are shown in the top insets of Figure 4c; these are to be compared with the angular dependence of the *z*- and *m*-modes in Figure 3c, confirming that peaks A and B have very different symmetry properties. We also performed similar polarization-resolved measurements on the $A_1$ and $E$ modes of Te powder as shown in the inset of the top panel in Figure 4b, which reflect the polycrystalline nature of the Te powder. The angular dependences of the two are similar, substantiating our speculation that the new peaks A and B in degraded *T'*-MoTe$_2$ are due to chemical reactions that generate clusters of Te metalloid.

To further understand the degradation rate of samples in different quality, we performed a set of controlled experiments using *in situ* Raman scattering to monitor sample degradation. Figures 5a and 5b show the time evolution of Raman spectra of a 'good' and a 'poor' sample. The spectra are recorded continuously in 30 seconds step for 5 minutes and the laser power is kept at 1mW focused to a spot size of about 1μm. In both samples, we observe increasing intensity of Te-like modes A and B as time goes but with different rates. For the 'good' sample, the intensity of Te-like modes grows while that of peaks due to 1L-*T'*-MoTe$_2$ decreases slowly. For the 'poor' sample, those 1L-*T'*-MoTe$_2$ modes disappear almost completely after the first 30 seconds; while the Te-like modes keep increasing and getting more and more intense, reaching more than twice the intensity of silicon peak in five minutes. As a controlled comparison, we also performed similar *in situ* Raman monitoring on a 'good' sample that was placed in the high vacuum pumped for a week at $3 \times 10^{-6}$ torr. In this sample, no additional Te-like modes were seen in 5 minutes, indicating negligible degradation. We summarize in Figure 5c the evolution of the intensity ratio of mode A to the silicon phonon at 520 cm$^{-1}$ for the three *in situ* Raman measurements. For the 'good' sample the Te(A)/Si ratio increases slowly; for the 'poor' sample, the Te(A)/Si ratio increases rapidly and saturates after 4 minutes. To further quantify the sample quality evolution, we extracted the linewidth and intensity of the 1L-*T'*-MoTe$_2$ $m^{85}$ mode in Fig.5a. The increase of FWHM from 3.4 to 4.6 cm$^{-1}$ during the 5 minutes indicates the degradation of sample crystallinity, correlated with the intensity decrease which is consistent with previous degradation studies of air sensitive 2D materials.[18,20,33]

In literature laser irradiation has been used to induce phase transition between *H*- and *T'*- MoTe$_2$ to make ohmic homojunction contacts,[23] where new Raman peaks at around 124 and 138 cm$^{-1}$ have been attributed to *T'*-MoTe$_2$ (similar conclusions have been reached on a study of strain induced *H* to *T'* phase transition).[22] We note that these peaks are quite similar to the Te metalloid Raman bands arising from sample degradation, raising questions regarding the final product of laser pattering and strain engineering in those samples.

**Conclusions**

In conclusion, the lattice dynamic of CVD 1L-*T'*-MoTe$_2$ is investigated by the polarization and crystal orientation resolved Raman spectroscopy. We observed the complete set of zone-center Raman-active modes including 3 *z*-modes and 6 *m*-modes, providing Raman fingerprints for high-quality 1L *T'*-MoTe$_2$. By monitoring the intensity of Raman features due to sample degradation, which are attributed to the Te-metalloid like phonons, we are able to quantitatively gauge the quality the 1L-*T'*-MoTe$_2$ crystal. Our work represents a solid advance in understanding the fundamental properties of *T'*-TMDC and provides a metrological tool for monitoring the quality of electrochemical and/or topological devices developed with *T'*-MoTe$_2$.

**Methods:**

**Crystal growth.**

Monolayer *T'*-MoTe$_2$ flakes were grown by chemical vapor deposition. We first spin coat a 1% sodium cholate solution on 300 nm SiO$_2$/Si substrates at 4000 rpm for 60 s. The molybdenum feedstock is provided by a droplet of a saturated solution of ammonium heptamolybdate in deionized water deposited onto the substrate. The prepared substrate is subsequently positioned in the center of the furnace, with 15 mg of solid tellurium placed 5 cm upstream from the substrate. At atmospheric pressure in a flow of 400 sccm of nitrogen gas and 25 sccm of hydrogen (both 99.999% purity), the furnace temperature is ramped at a rate of 70 °C min$^{-1}$ to 700 °C. Meanwhile the temperature of the tellurium pellet is around 500 °C. After a 5 min growth period, the sample is rapidly cooled down to room temperature by flowing 1000 sccm nitrogen gas with the furnace lid open. Bulk *T'*-MoTe$_2$ crystals are grown *via* chemical vapor transport (CVT) method using bromine as the transport gas. The source materials, including Mo(99.9%), Te(99.997%), and TeBr$_4$(99.999%) powders, are placed in a fused silica tube. Total amount of Mo and Te are kept in a stoichiometric 2:1 ratio with sufficient TeBr$_4$ to achieve a Br density of 3 mg/cm$^3$. The tube is pump-purged with argon gas (99.999%) and sealed at low pressure prior to growth. The CVT growth is performed using a three-zone furnace, with the temperature setting of the reaction and growth zones at 1000 and 900 °C, respectively, for 100 h. At the end of the growth, the crystal is thermally quenched by a water bath.

**Sample passivation.**

The graphene for passivating 1L-$T'$-MoTe$_2$ is firstly grown on a Cu foil substrate by CVD. It was then transferred off the Cu foil substrate by bubble transfer method with a NaOH solution and left afloat in a DI water bath. Following the growth of 1L-$T'$-MoTe$_2$ by CVD, the sample is briefly dipped inside the DI water bath and the graphene is instantly pulled over the chip to cover the 1L-$T'$-MoTe$_2$. The graphene/1L-$T'$-MoTe$_2$ stack is then immediately dried with N$_2$ gun. Through this quick passivation method, the 1L-$T'$-MoTe$_2$ flakes are in air and water for a handful of seconds which minimizes the degradation.

**Transmission Electron Microscopy.**

The as grown 1L-$T'$-MoTe$_2$ flakes were transferred onto a commercial holey-carbon TEM grid using a typical wet transfer method. The TEM was performed with a JEOL ARM 200CF equipped with a CEOS corrector and a high-angle annular dark field detector. The operation voltage is kept below 80 kV to avoid sample damage. Selected-area electron diffraction patterns were acquired in TEM mode using an aperture with an effective size of ~1 μm at the sample.

**Acknowledgement**

This work is supported by the University of Massachusetts Amherst and in part by the Armstrong Fund for Science. C.H.N. and A.T.C.J. acknowledge the support of the National Science Foundation MRSEC program through grant number DMR-1120901. The authors would like to thank William M. Parkin from the Drndić Group at the University of Pennsylvania for the TEM and selected-area electron diffraction images of the $T'$-MoTe$_2$ flakes.

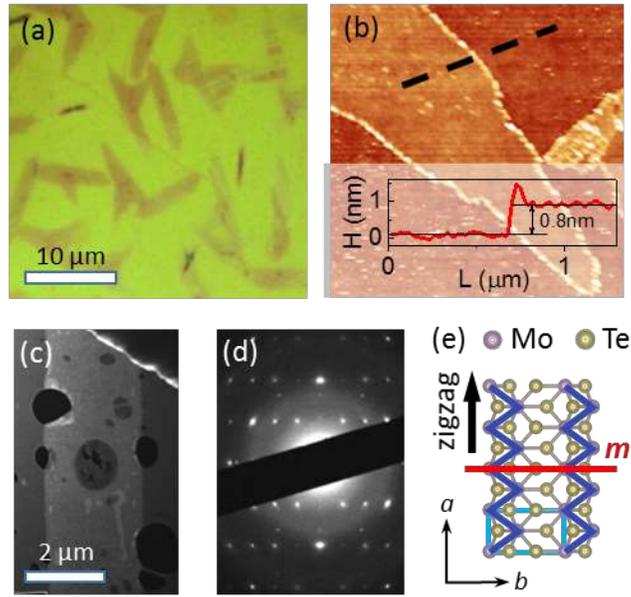

**Figure 1.** (a) The optical microscope (OM) image of a typical sample. All bamboo-leaf like flakes on the image are 1L-*T'*-MoTe$_2$, and the small dark dots near center of the flakes are multilayer crystals. (b) AFM image of a monolayer flake. The step height is 0.8 nm. (c) Dark field TEM image of a 1L-*T'*-MoTe$_2$ flake transferred on top of holey carbon film. (d) The selected-area electron diffraction image of a suspended 1L-*T'*-MoTe$_2$, exhibiting rectangular diffraction patterns. (e) Schematic top view of the 1L-*T'*-MoTe$_2$ crystal. The *a*-axis is aligned with the zigzag direction. The Mo-Mo zigzag chains are highlighted by blue lines. The mirror plane which is perpendicular to zigzag chains is shown as a red line. The unit cell is denoted as a light-blue rectangle.

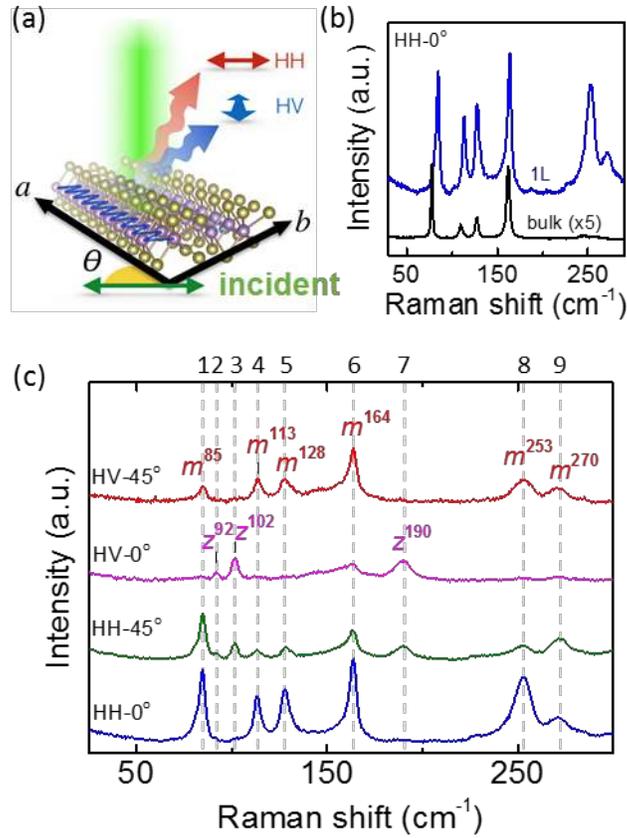

**Figure 2.** (a) Schematics of light-polarization and crystal-orientation resolved Raman spectroscopy. The polarization of incident light (green arrow) is set at an angle $\theta$ with respect to the *a*-axis of the crystal. The scattered light is selectively collected in HH or HV configurations, corresponding to the detection of photons with co-linear or cross-linear polarization, respectively. (b) Typical Raman spectra of 1L and bulk *T'*-MoTe$_2$ with $\theta = 0°$ in HH configuration. (c) The Raman spectra of 1L-*T'*-MoTe$_2$ detected with different combination of incident polarization angles and collection geometries. In HV configuration, the phonon modes are highly selective, showing 6 *m*-modes ($m^{85}$, $m^{113}$, $m^{128}$, $m^{164}$, $m^{253}$, $m^{270}$) at $\theta = 45°$ and 3 z-modes ($z^{92}$, $z^{102}$, $z^{190}$) at $\theta = 0°$.

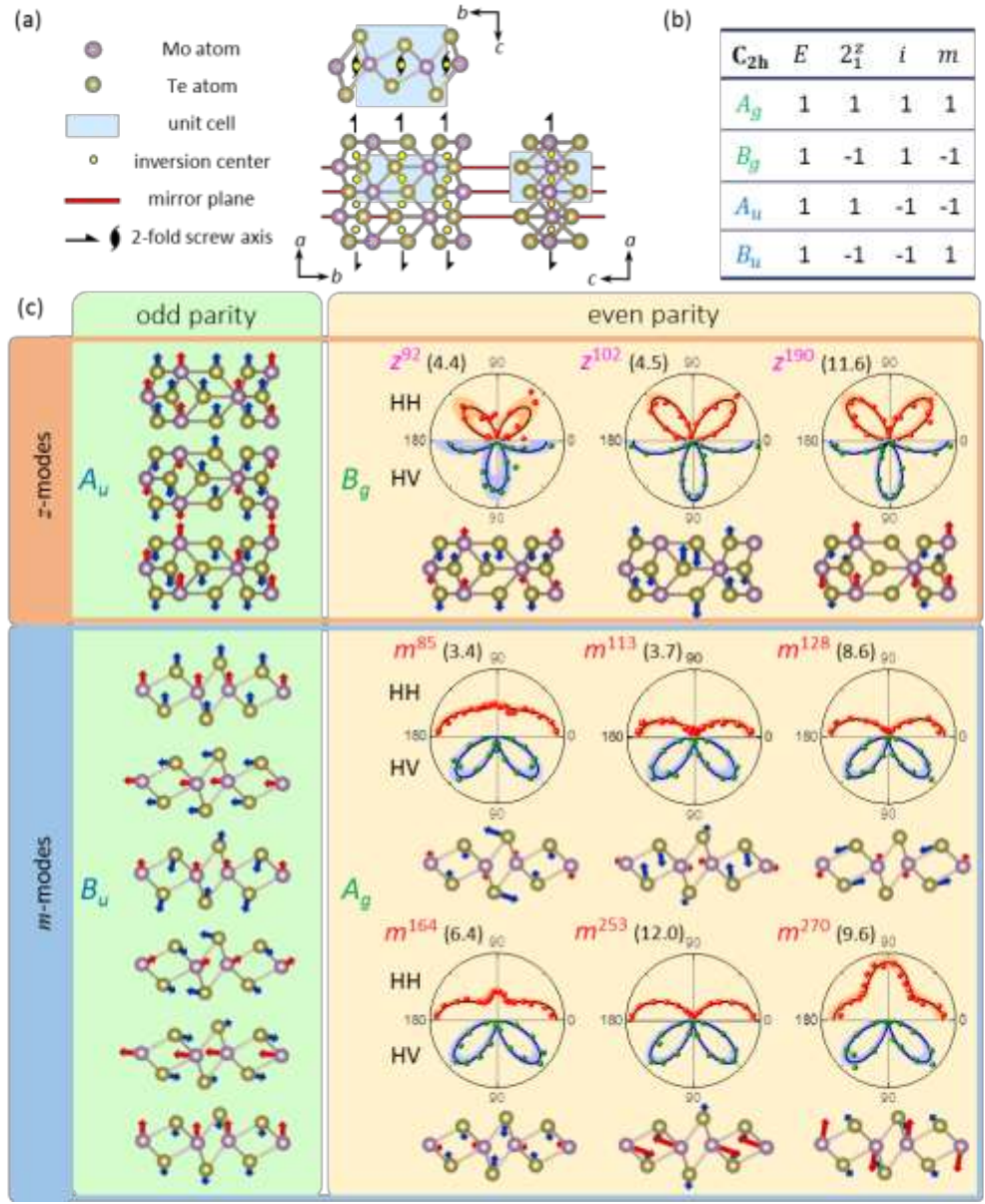

**Figure 3.** (a) Top and side views of 1L-$T'$-MoTe$_2$ atomic arrangement. The unit cell and the symmetry operations are illustrated on top of the schematic drawings. (b) Character table of the $C_{2h}$ group. (c) The schematics of all zone-center normal modes categorized into 4 groups with different symmetry: $z$-modes with odd parity belong to $A_u$ symmetry; $z$-modes with even parity belong to $B_g$ symmetry; $m$-modes with odd parity belong to $B_u$ symmetry; and $m$-modes with even parity belong to $A_g$ symmetry. The intensity angular dependences of the 9 Raman-active modes are plotted above the corresponding lattice vibrations. The FWHM of each mode is included in the parentheses.

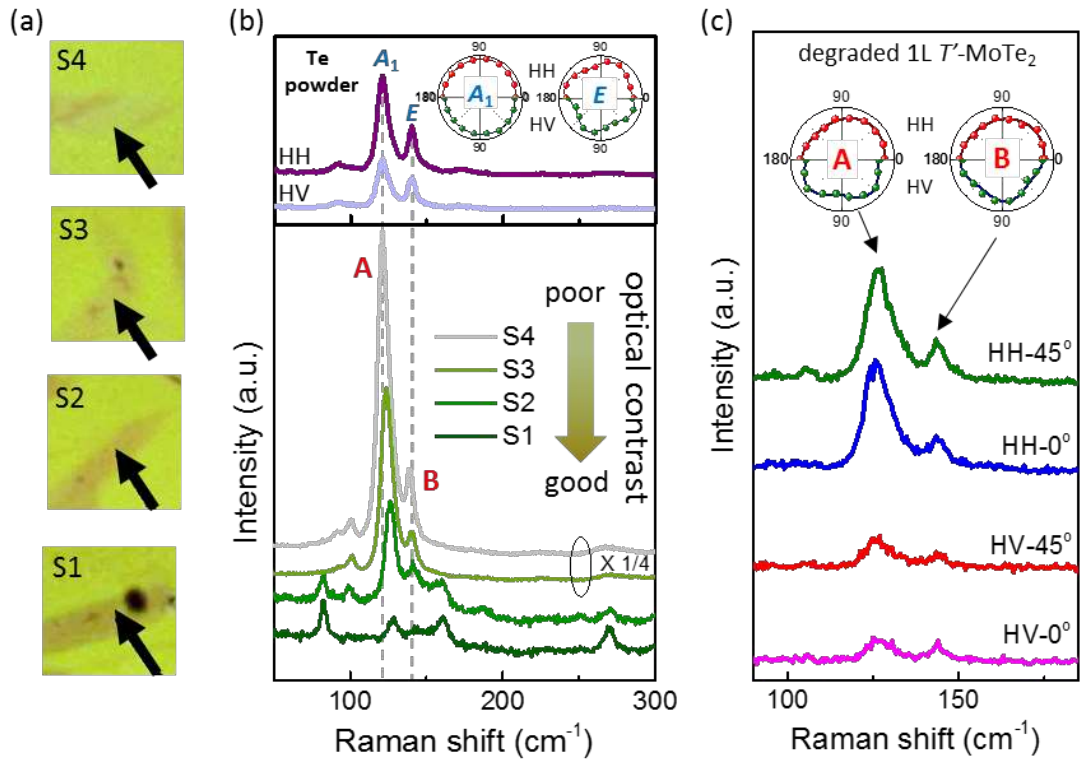

**Figure 4.** (a) OM images of selected 1L-$T'$-MoTe$_2$ with different optical contrast: S1 to S4 from good to poor contrast. The arrows point to the positions where Raman spectra were collected. (b) Lower panel: the Raman spectra for samples S1 to S4. The spectra are shifted vertically for clarity. The Raman spectra from Te powder are plotted for comparison in the upper panel. Insets show detailed angular dependence of Raman intensity in HH and HV for two prominent peaks $A_1$ and $E$ of Te powder. (c) The Raman spectra of the sample with poor contrast in HH or HV configuration with $\theta = 0°$ and 45°. The angular dependences of intensity for peaks A and B of degraded 1L-$T'$-MoTe$_2$ are shown in the inset.

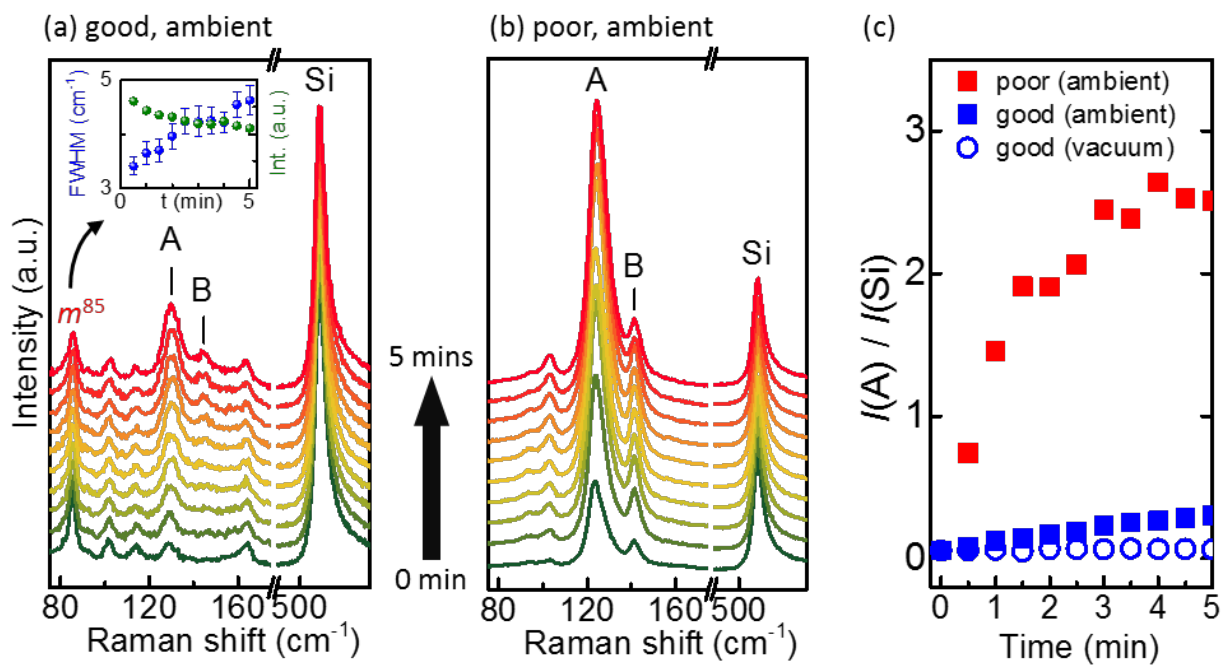

**Figure 5.** (a)&(b) Spectral evolution of samples with (a) good (similar to S1 in Figure 4a) and (b) poor (similar to S4 in Figure 4a) optical contrast in ambient. The inset in panel (a) shows the evolution of FWHM and intensity of the $m^{85}$ mode. (c) The intensity ratio of peak A to the silicon mode plotted as a function of laser exposure time.

**S1. Crystal Orientation Resolved Raman Scattering on polycrystalline 1L-*T'*-MoTe₂.**

Here, we demonstrate that the angular dependence of intrinsic Raman modes of 1L-*T'*-MoTe$_2$ can be used to determine the crystal orientation of each domain in a polycrystal. We select a polycrystalline 1L *T'*-MoTe$_2$ which grows in two different directions in two regions, as shown in Figure S1(a). The crystal orientation resolved Raman scattering is performed on the two spots indicated as a red and a green dot in HH geometry. We then extract the angular dependent intensity for two intense Raman bands, $m^{164}$ and $m^{253}$. The fitted direction of crystal orientation (*a*-axis) are indicated by arrows in the center of the polar plots in Figure S1(b) and S1(c). The difference of crystal orientation between the two pieces is 60.0° ± 0.5°.

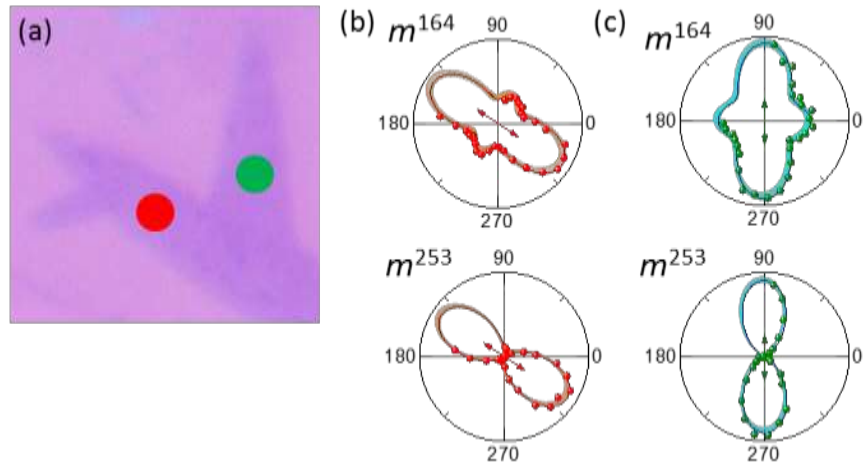

**Figure S1.** (a) OM image of a selected polycrystalline 1L-*T'*-MoTe$_2$ flake. (b)&(c) The crystal orientation dependent intensity of mode $m^{164}$ and mode $m^{253}$. (b) is for crystal on the left (red spot on Figure S1a) and (c) is for crystal on the right (green spot on Figure S1a). It is clear that the two leaves have different crystal axis directions. The arrows in the center of angular patterns indicate the fitted direction of the zigzag chain (*a*-axis).